# Gerberto e le fistulae: tubi acustici ed astronomici

**Costantino Sigismondi[1]**

[1]ICRA/Sapienza Università di Roma e Pontificia Università Regina Apostolorum, sigismondi@icra.it

**Abstract.** *Gerbert and the fistulae: acoustical and astronomical tubes*

*Gerbert of Aurillac wrote to Constantine of Fleury in 978 a letter to describe in detail the procedure to point the star nearest to the North celestial pole. This was made to align a sphere equipped with tubes to observe the celestial pole, the polar circles, the solstices and equinoxes. The use of tubes in astronomical observation is later reported by Alhazen in his treatise on optics (1011-1021).*

*The description of pointing to the celestial pole indicates that the instrument must be accurately aligned with the true pole, materialized at that epoch by a star of fifth magnitude, at the limit of naked eye visibility, and then the instrument must remain fixed.*

*Solstices and Equinoxes are points of the orbit of the Sun, so the sphere could be used as a tool for observing the Sun and probably determine the duration of the tropical year. This sphere was much more than a didactic tool, given the long procedure for the accurate alignement.*

*Moreover "Rogatus a pluribus" (asked by his many students), Gerbert wrote a treatise on acoustic tubes (fistulae) in 980: Mensura Fistularum.*

*He knew the difference in behavior of the fistulae compared with the acoustic strings, already studied by the Pythagoreans, and the treaty is intended to present the law that governs the length of the organ pipes in two octaves, compared to the corresponding acoustic strings. In terms of modern physics we know that acoustic tubes require an "end correction" to be tuned, which is proportional to the diameter of the tube. This proportionality is the same for every note. The mathematical law is simple, but Gerbert preferred to create a law in which the proportions of pipes and strings should be calculated through a series of fractions linked to the number 12 and its multiples.*

## 1. Introduzione

Gerberto nato attorno al 945, molto probabilmente da nobile stirpe,[1] fu monaco benedettino ad Aurillac. Nel 967 partì, con speciale dispensa, alla volta della Catalogna per studiare matematica presso l'arcivescovo Attone di Vich. Nel 971, già famoso per la sua conoscenza in musica,[2] era a Roma presso il papa Giovanni XIII e nel 972 seguiva Geranno di Reims nella scuola cattedrale più importante di Francia, di cui presto divenne direttore oltre che segretario dell'arcivescovo Adalberone di Reims.

Tra il 980 ed il 984 fu anche abate a Bobbio, dov'era la biblioteca più grande d'Europa. Gerberto sostenne la politica di Adalberone pro Capetingi ed a favore del Sacro Romano Impero degli Ottoni, ma quando nel 989 il nuovo re di Francia Ugo Capeto designò il successore di Adalberone, la scelta cadde sul carolingio Arnolfo. Due anni dopo un concilio di vescovi lo depose eleggendo Gerberto al suo posto, ma le disposizioni di questo concilio non furono mai accettate dal papa e Gerberto si ritiro' di buon ordine a Saasbach presso la corte del giovanissimo Ottone III di cui era stato docente. Nominato nel 998 arcivescovo di Ravenna Gerberto fu eletto papa il 2 aprile 999 a seguito dell'improvvisa morte di Gregorio V cugino dell'imperatore regnante. Da pontefice prese il nome di Silvestro II ed allargò i confini giuridici della Chiesa alla Polonia e all'Ungheria ed istituì la commemorazione dei defunti il 2 novembre. Morì il 12 maggio 1003. Il suo prestigio culturale di livello europeo gli consentì di rinforzare l'immagine del papato e della Cattedrale di S. Giovanni in Laterano come sua sede, e a tutt'oggi lì si può vedere il suo epitaffio tombale, risistemato sul secondo pilastro della navata destra da Francesco Borromini nel 1648.

---

[1] Flavio G. Nuvolone ha avanzato recentemente l'ipotesi che Gerberto discendesse dalla famiglia Carlat, cfr. NUVOLONE, F. G. *Il numero e la croce, l'Homo Novus da Aurillac*, Nuovo Medioevo **87**, Liguori, Napoli (2012).

[2] SANTI, E. *Gerberto e la musica*, Geografia 103-104, 81 (2003).

La maggior parte dei suoi scritti è giunta fino a noi raccolta nel volume 139 della Patrologia Latina dove troviamo trattati sull'astrolabio, sulla geometria, sulla filosofia degli universali, ed un epistolario di oltre 200 lettere. Altri scritti matematici[3] e musicali sono stati raccolti da vari studiosi; la sua biografia fino all'anno 999 scritta da Richero di Reims, suo allievo, è nel volume 138 della stessa Patrologia.

Gerberto è stato il personaggio più rappresentativo del X secolo in Europa, tramite il quale si possono conoscere molti aspetti della vita politica, culturale e religiosa del tempo. Dopo la sua morte e per almeno sei secoli Gerberto non è stato compreso adeguatamente e molte leggende sono nate attorno a lui, che avrebbe conseguito tanta scienza per intervento del diavolo. Oggi la sua figura è oggetto di nuovi studi che gettano una nuova luce anche sulla scienza nel periodo a cavallo tra alto e basso medioevo.

Qui esaminiamo uno strumento astronomico progettato da Gerberto e descritto nella lettera a Costantino di Fleury nel 978, e l'acustica delle canne d'organo. Gerberto fece costruire a Bobbio un organo per omaggiare Ottone II nel 980. L'organo era uno strumento allora veramente raro in Europa, e Gerberto scrisse anche un trattato teorico sulle canne d'organo nello stesso anno.

## 2. La sfera astronomica

Gerberto scrisse a Costantino di Fleury nel 978 una lettera in cui descriveva accuratamente la procedura di puntamento con un tubo della stella che individuava, a quel tempo, il polo Nord celeste, per allineare una semisfera munita di tubi per osservare poli, circoli polari, solstizi ed equinozi. L'uso di tubi nell'osservazione astronomica[4] richiama, a noi moderni, il telescopio, ma già Alhazen[5] all'inizio del XI

---

[3] OLLERIS, A. *Oeuvres de Gerbert, Pape sous le nom de Silvestre II*, Thibaud, Clermont-Ferrand (1867).
   BUBNOV, N., *Gerberti Opera Mathematica*, Berlin (1899) reprint Hildesheim (1963).

[4] CAVICCHI, E. *Activity inspired by Medieval Observers and Tubes*, in Orbe Novus, Universitalia Roma, 23, (2010).

[5] ALHAZEN IBN AL-HAYTHAM, *Kitab al-Manazir (Opticae Thesaurus Alhazen Arabis; De Aspectibus; Perspectiva)*, (1011-1021).
   SABRA, A. I. ed. (1983) *The Optics of Ibn al-Haytham, Books I-II-III: on direct vision*; (2002) *Books IV-V: on reflection and Images seen by reflection*, Kuwait: National Council for Culture, Arts secolo ne parlava nel suo trattato di ottica. Nella sfera di Gerberto, antecedente di oltre trent'anni al trattato di Alhazen, ce ne sono almeno cinque.

La descrizione del puntamento al polo celeste indica che lo strumento doveva essere allineato accuratamente con il vero polo, materializzato da una stella di quinta grandezza, al limite della visibilità ad occhio nudo. La laboriosità della procedura, che richiedeva anche più notti di osservazione, ha come logica conseguenza che poi lo strumento doveva rimanere fisso.[6]

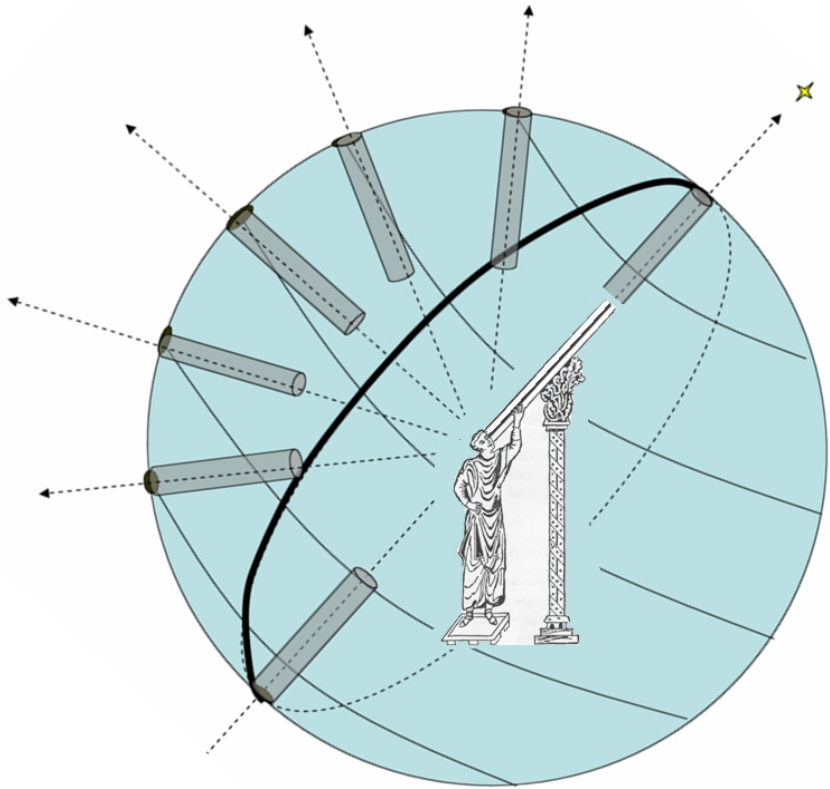

*Fig. 1 - Ricostruzione della Sfera di Gerberto, utlizzando un'immagine di un manoscritto di San Gallo del 982 (ora perduto) che rappresenta un uomo che osserva con un tubo ed il testo della lettera di Gerberto a Costantino di Fleury. La rotazione della semisfera ricopre tutta la sfera e consente l'osservazione di tutti i cerchi paralleli dei circoli polari, tropici ed equatori più i due poli.*

Solstizi ed equinozi sono luoghi dell'orbita del Sole, dunque la sfera poteva essere uno strumento per osservare il Sole e probabilmente proprio determinare la durata dell'anno tropico.

La sfera era molto più che uno strumento per uso meramente didattico o dimostrativo, data la procedura accurata di puntamento. Se il suo scopo fosse stato dimostrativo era inutile puntarlo esattamente. La stella polare al tempo di Gerberto era la HR4893 di

---

and Letters.
[6] Pratt Lattin, H. *The Letters of Gerbert,* Columbia University Press, New York, 1961.

magnitudine 5.28 che a quel tempo veniva chiamata *Polus*. Oggi è anche chiamata 32 Camelopardalis. Essa è ben distinta dall'attuale stella Polare, che al tempo si chiamava *Computatrix*, come si vede bene nei disegni in Fig. 2, 3 e 4.

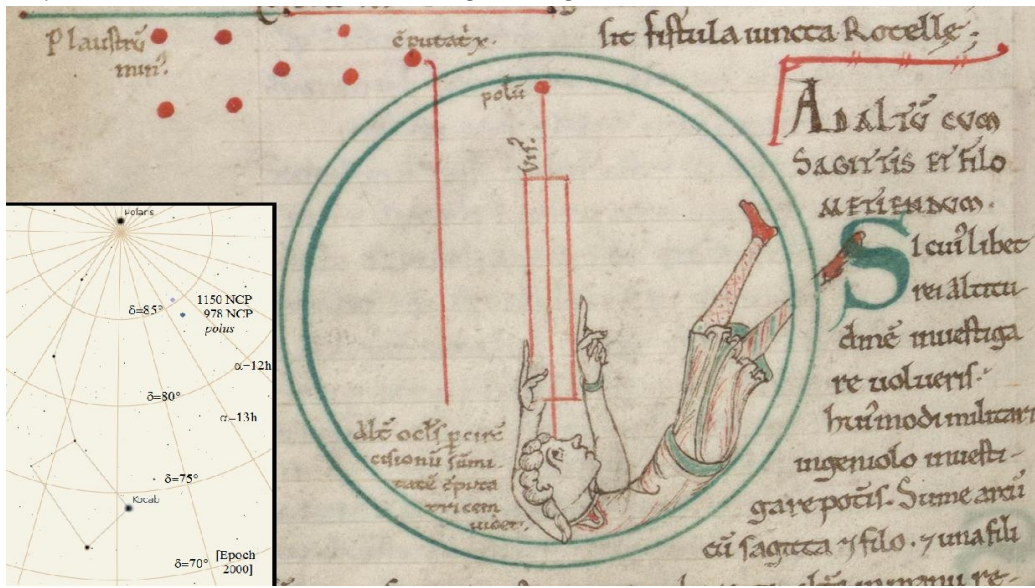

*Fig. 2 - Disegno tratto dal manoscritto di Avranches BN 235, f. 32, XII secolo. L'uso della fistula per puntare la stella chiamata Polus mentre l'altro occhio punta un'altra stella, la Computatrix, che per noi del XXI secolo è l'attuale Polare. La costellazione che è rappresentata è l'Orsa Minore, altrimenti chiamata Cynosura o Plaustrum minus (come nel mondo anglosassone il "piccolo mestolo"). Nel riquadro è indicata la posizione del polo Nord celeste nel 978 e nel 1150 e quella della stella Polus.*

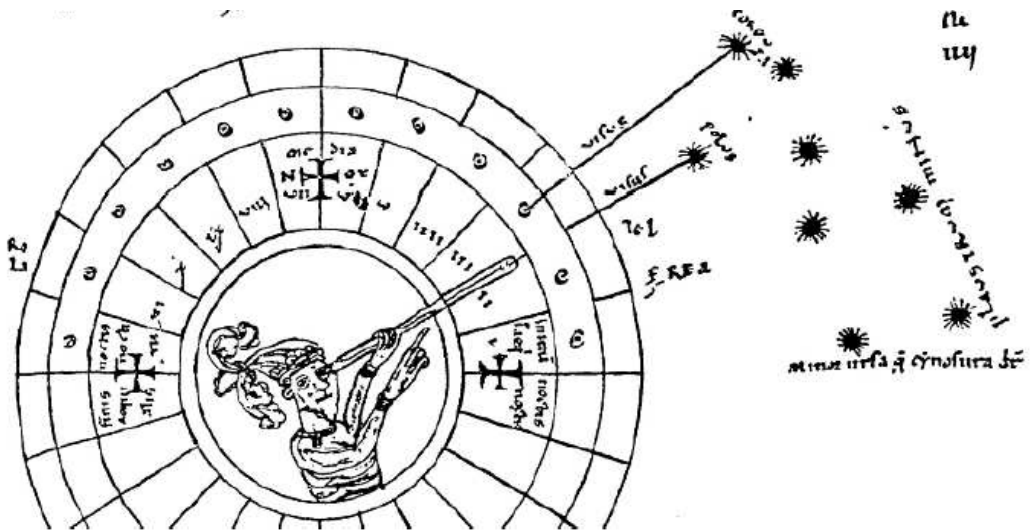

*Fig. 3 - Riproduzione dal manoscritto 173 di Chartres, distrutto durante un bombardamento della seconda guerra mondiale. Viene mostrata la stessa mappa stellare con chiara indicazione delle stelle Polus e Computatrix.*

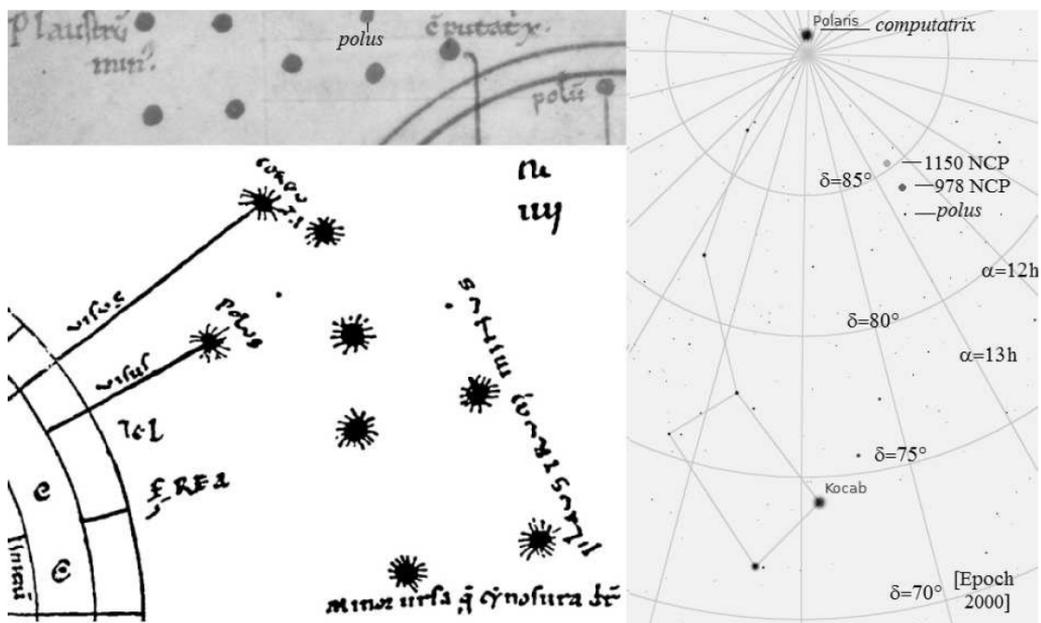

*Fig. 4 - Riproduzione di parte del disegno del manoscritto 173 di Chartres a confronto con la parte di mappa stellare nel manoscritto di Avranches BN235 f. 32, e mappa stellare attuale con l'indicazione della stella Polus e della Computatrix. A destra è indicata la posizione del polo Nord celeste (NCP) nel 978 e nel 1150 e quella della stella Polus.*

La stella *Polus* è osservabile ad occhio nudo sotto un cielo buio. L'autore la ha osservata dal finestrino dell'aereo in viaggio per la Cina la notte tra l'8 ed il 9 ottobre 2011, durante la pioggia di stelle cadenti delle Draconidi.[7]

Il vescovo Tietmaro di Merseburgo[8] (975-1018) riportò nel suo Chronicon per l'anno 1014 di un *"horalogium"* lasciato da Gerberto in Magdeburgo: "In Magdeburg horalogium fecit, recte illud constituens secundum quandam stellam, nautarum ducem, quam consideravit per fistulam miro modo" ed ancora visibile al tempo in cui Tietmaro scriveva. Questo strumento permanente poteva essere quello descritto a Costantino di Fleury nel 978, oppure un notturlabio[9] fisso e grande abbastanza per essere visibile da tutti i visitatori di Magdeburgo, come riportato da Tietmaro.

Una volta allineato lo strumento col polo Nord celeste sarebbe servita solo un'altra stella per computare l'ora della notte in una certa data dell'anno. Si noti come la linea tirata tra la stella *Polus* e la *Computatrix* - attuale Polare conteneva sia il polo Nord celeste del 978 che quello nel 1150, data del manoscritto su cui è riportato il disegno.

---

[7] SIGISMONDI, C. *Airborne observation of 2011 Draconids meteor outburst: the Italian mission*, 2011 http://arxiv.org/pdf/1112.4873v1

[8] THIETMARI MERSEBURGENSIS EPISCOPI, *Chronicon*, VI 100 (1018).

[9] OESTMANN G. *On the history of the nocturnal*, Bulletin of the Scientific Instrument Society, 69, 5 (2001).

Questo fatto rende il notturlabio usato da Gerberto molto precisio per determinare l'ora della notte.

Probabilmente dobbiamo proprio all'*horologium* messo a punto da Gerberto a far sì che l'attuale stella Polare fosse chiamata *Computatrix* data la sua funzione per calcolare l'ora.[10]

Un tubo fisso creato da Nicola Cusano quasi cinque secoli dopo Gerberto, potrebbe essere stato usato per determinare l'istante del solstizio invernale;[11] questo mostra come l'astronomia prima del telescopio potrebbe ancora rivelarci delle sorprese, anche se dobbiamo tenere presente sempre il contesto scientifico e culturale dell'epoca prima di arrivare a conclusioni mirabolanti.

### 3. Le canne d'organo

*Rogatus a pluribus*, dai suoi studenti, Gerberto scrisse il trattato sulle fistulae acustiche nel 980, la *Mensura Fistularum*.[12]

Egli conosceva la differenza di comportamento delle fistulae, dei tubi acustici, rispetto alle corde acustiche, già studiate dai pitagorici, ed il trattato è finalizzato a presentare le leggi che governano la lunghezza delle canne d'organo in due ottave, rispetto alle corrispondenti corde acustiche.

Nella fisica moderna i tubi acustici, per essere intonati, necessitano di una «end correction» una correzione di bocca, proporzionale al diametro del tubo. Questa proporzionalità è la stessa per ogni nota. La legge matematica è semplice, ma Gerberto preferì creare una legge in cui le proporzioni tra canne e corde dovessero essere calcolate mediante una serie di frazioni riconducibili al numero 12 e ai suoi multipli.[13]

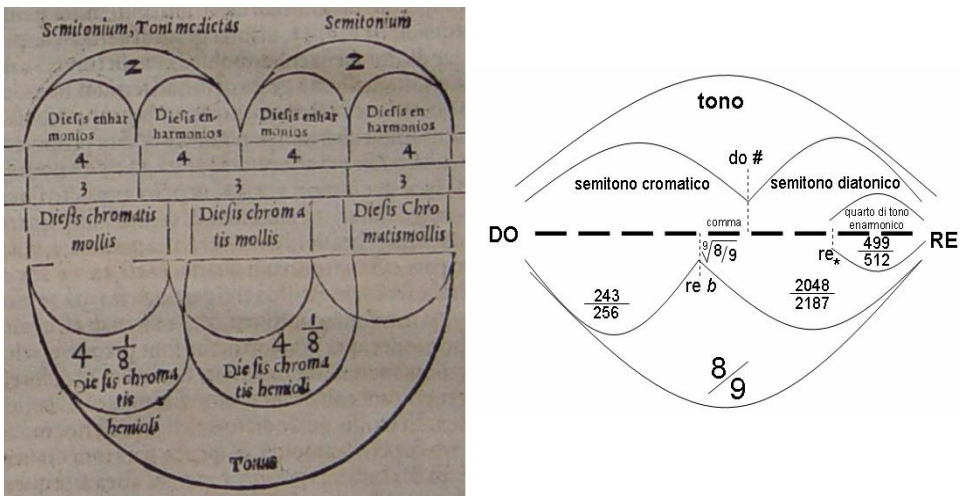

*Fig. 5 - Suddivisione dell'intervallo di un tono secondo Aristosseno (sinistra) e in una visione sinottica della musica antica.*[12]

---

[10] MICHEL, H. *Les tubes optiques avant le télescope*, dans Ciel et terre, bulletin de la Societé belge d'astronomie, de méteorologie et de physique du globe 70, 175 (1954).

[11] DE DONÀ, G. *Nicola Cusano e il foro astronomico al castello di Andraz*, Gerbertus 1, 251 (2010).

[12] SACHS, K -J. *Mensura Fistularum*, Stuttgart-Murrhardt, 2 vol., (1970-1980).

[13] SIGISMONDI, C. *Gerberto e la misura delle canne d'organo*, Archivum Bobiense 29, 355 (2007).

Era più importante per Gerberto mostrare che l'acustica, la musica, ubbidiva a delle leggi matematiche che ricordavano l'importanza del numero 12, tanto nell'Antico quanto nel Nuovo Testamento, piuttosto che rappresentare il fenomeno in un modo che per noi moderni sarebbe stato più semplice e compatto, identico tanto per la prima quanto per la seconda ed ogni eventuale altra ottava. Le canne d'organo Gerberto le faceva costruire proprio nella fonderia di Bobbio, dove fu abate dal 980 al 984. Il dono a Ottone II, per cui aveva preparato anche un libretto con un Carme Figurato, era stato preparato nel 980 che è la stessa data di pubblicazione della *Mensura Fistularum*. In quel Carme, al secondo livello di criptazione, come era in uso nelle élites intellettuali dell'epoca, Gerberto lascia la traccia dei numeri arabi incluso lo zero.[14] È la prima volta nell'occidente cristiano. Egli aveva già introdotto l'abaco nella sua scuola di Reims che dirigeva, usando 9 cifre e lasciando un posto vuoto al posto dello zero. Di Gerberto musico si conosce anche che egli compose un inno allo Spirito Santo, ma non ne è stata ancora trovata alcuna traccia.

**4. Conclusioni**

La divisione del sapere in trivio e quadrivio, già fatta da Boezio, continuò nella scuola cattedrale di Reims dove Gerberto rinforzò molto il quadrivio, Matematica, Geometria, Musica e Astronomia, con metodi di insegnamento sperimentali, ricordati con entusiasmo dal suo primo biografo, Richero di Reims. Il ruolo dell'insegnamento delle materie del quadrivio nelle future Università trova in Gerberto, poi divenuto papa Silvestro II, il suo ispiratore più autorevole. Gerberto realizzò anche un abaco,[15] e coloro che lo usavano venivano chiamati *gibertisti* fino al XIV secolo. Già lui era capace di effettuare calcoli complessi con rapidità e precisione, raggiungendo una precisione molte volte maggiore dei testi usati in geometria al suo tempo, ad esempio nel calcolo dell'area del triangolo equilatero Gerberto propose un metodo che approssimava il valore al meglio dell'1%, mentre il precedente dava un errore del 33%.[16] Introdusse anche i numeri arabi nell'Europa cristiana, ma la loro conoscenza restò circoscritta ad una ristretta cerchia intellettuale. È di Gerberto anche il trattato sull'astrolabio, strumento che lui conobbe molto probabilmente durante il suo soggiorno in Catalogna. Le sfere gerbertiane sono comunque una rielaborazione originale sia del sapere astronomico arabo del tempo sia di quello latino.[17]

La stella di Gerberto, chiamata *polus* durante tutto l'alto medioevo, ancora è muta testimone del primo strumento a montatura equatoriale realizzato da Gerberto; dobbiamo attendere il padre gesuita Chirstoph Grienberger (1561-1636) negli anni seguenti al 1610 per avere il progetto della prima montatura equatoriale usata per le osservazioni del Sole al telescopio.[18]

---

[14] Nuvolone, F. G. in *Culmina Romulea*, 58, Ateneo Pontificio Regina Apostolorum Roma (2008).

[15] Falcolini, C. *Gerberto e l'abaco*, in Doctissima Virgo, 113, APRA Roma (2011).

[16] Rossi, P. *Algoritmi matematici nelle lettere di Gerberto*, Gerbertus 1, 16 (2010).
Sigismondi, C. *Gerbert of Aurillac: astronomy and geometry in tenth century Europe*, Nuovo Cimento B in stampa, http://arxiv.org/abs/1201.6094 (2012).

[17] Zuccato, M. *Gerbert's Islamicate Celestial Globe*, Archivum Bobiense Studia V, 167 (2005).

[18] Secchi, A. *Il Sole*, Tipografia della Pia Casa di Patronato, Firenze (1884).